\documentclass[twocolumn,prb,showpacs,floatfix, ]{revtex4}  
\usepackage{color,graphicx}
\usepackage{verbatim}
\usepackage{amssymb}

\usepackage{latexsym}
\definecolor{red}{rgb}{0.8,0,0.2}

\def\beeq{\begin{equation}}
\def\eneq{\end{equation}}
\def\beeqa{\begin{eqnarray}}
\def\eneqa{\end{eqnarray}}

\begin{document}
\DeclareGraphicsExtensions{.ps,.pdf,.eps}

\title{Charge-dependent migration pathways for the Ga vacancy in GaAs}

\author{Fedwa El-Mellouhi} \email {f.el.mellouhi@umontreal.ca}
 \affiliation{D\'epartement de physique and Regroupement qu\'eb\'ecois
   sur les mat\'eriaux de pointe, Universit\'e de Montr\'eal, C.P.
   6128, succ. Centre-ville, Montr\'eal (Qu\'ebec) H3C 3J7, Canada}

\author{Normand Mousseau } \email {Normand.Mousseau@umontreal.ca}
 \affiliation{D\'epartement de physique and Regroupement qu\'eb\'ecois
   sur les mat\'eriaux de pointe, Universit\'e de Montr\'eal, C.P.
   6128, succ. Centre-ville, Montr\'eal (Qu\'ebec) H3C 3J7,
   Canada\footnote{permanent address}}

\affiliation{Service de Recherches de M\'etallurgie Physique, Commissariat ˆ
  l'\'energie atomique-Saclay, 91191 Gif-sur-Yvette, France.}

\date{\today}
\begin{abstract}

{Using SIEST-ART, a combination of the local-basis \textit{ab-initio}
program SIESTA and the activation-relaxation technique (ART nouveau)
we study the diffusion mechanisms of the gallium vacancy in
GaAs. Vacancies are found to diffuse to the second neighbor using two
different mechanisms, as well as to the first and fourth neighbors
following various mechanisms. We find that the height of the energy
barrier is sensitive to the Fermi-level and generally increases with
the charge state. Migration pathways themselves can be strongly
charge-dependent and may appear or disappear as a function of the
charge state. These differences in transition state and migration
barrier are explained by the charge transfer that takes place during
the vacancy migration.  }
\end{abstract}
\date{\today}

\maketitle

\section {Introduction} 

Self-diffusion is one of the basic mass-transport mechanisms in materials. While it is
one of the most powerful tools used in the preparation of
nanostructures~\cite{3Wan04, 3Wan05, 3Ras04,3Yu05_2}, many questions remain
regarding the microscopic details of self-diffusion and recent studies have
shown that even in the simplest cases, the mechanisms can be much more
complicated than was initially thought. Diffusion on simple metallic surfaces,
for example, was found to take place through a range of mechanisms involving
from one to at least seven atoms~\cite{3Fei90,3Hen00}. Similarly, recent studies
on self-interstitial clusters~\cite{3Cog05, 3Du05} and disordered
systems~\cite{3Mou98,3wales} have highlighted the importance of collective moves
in easing atomic motion even in bulk systems and underlined the importance for
a detailed characterization of these mechanisms in materials of technological
interests. This is the case for semiconductors, for example, that are at the
heart of the electronic industry. While one predicts, using symmetric
considerations, the self-diffusion pathways in elemental materials, such as
silicon, this approach becomes impossible when impurities are added or
multi-compound systems are considered. Thus, one must turn to experiments or
numerical simulations to provide a direct and comprehensive study of diffusion
mechanisms in semiconductors. The difficulty to extract precise information on
the diffusion mechanisms in these materials is compounded by the nature of
bonding and the importance of charged defects that complicate seriously both
experimental and theoretical studies.

Previous numerical studies of the migration pathways of intrinsic
defects in binary semiconductors have focused on GaAs~\cite{3Boc96},
SiC~\cite{3Boc03,3Rur03} and GaN~\cite{3Lim04}. Except in one
case~\cite{3Boc96}, where high-temperature molecular dynamics was used,
these works focus on optimizing preselected pathways using
algorithms such as the drag~\cite{3Boc96} or the ridge
method~\cite{3Ion93} that rely on the knowledge of the initial and
final states in addition to a decent guess of the overall diffusion
mechanism. While these approaches work efficiently to identify with
precision the migration energy of previously known diffusion
trajectories, they cannot help identify complex or unexpected
mechanisms that could also play an important role in the diffusion
process. 

Here, we present the application of the activation-relaxation
technique in its more recent implementation (ART nouveau) to explore
systematically the diffusion mechanisms of complex systems. More
precisely, we focus on the diffusion of $V_{Ga}$ in GaAs because of
its apparent simplicity but also because of its technological interest
and its role in affecting the properties of bulk materials and
nanostructures. Gallium vacancies are found to be mobile at typical
growth and annealing temperatures~\cite{3Cor92}--- with a dominant charge state
strongly depending on growth conditions, temperature, dopants,
etc.--- playing the main role in dopant diffusion.  For
example, Bracht {\it et al.}~\cite{3Bra05} showed recently that the
contribution of $V_{Ga}$ to Ga self-diffusion in GaAs is even more
important than earlier estimation giving an important contribution to
the total diffusion profile.  Furthermore, Tuomisto {\it et
al.}~\cite{3Tuo04} found that Ga vacancies play a central role in the
migration of Mn in Ga$_{\text{1-x}}$Mn$_{\text{x}}$As alloys. Finally,
the energy blueshift in PL spectra of InGaAs/GaAs~\cite{3Dji06} and
InAs/GaInP quantum dots~\cite{3Raz04} has also been recently attributed to
$V_{Ga}$ diffusion.

Focusing on a simple defect, a gallium vacancy ($V_{Ga}$), in the
weakly ionic GaAs, we show here that diffusion in bulk semiconductors
is a complex phenomenon that depends not only on the geometry of the
defect and the surrounding lattice but also on  its charge. In particular, we identify a new mechanism for the diffusion
to the second neighbor in addition to the one already found by
Bockstedte and Scheffler~\cite{3Boc96}, plus a number of other jumps to
the first and fourth neighbors.  Not all these pathways are likely to
occur in a normal range of temperature, and some exist only for a
subset of charge states, but their existence underlines the
under-estimated richness of diffusion mechanisms in bulk materials.

This paper is organized as follows, in Sec.~\ref{detail}, we present the
details of activated event generation using SIEST-A-RT. A description of
initial and different transition states in various charge states are presented
in Sec.~\ref{results}. In Sec.~\ref{discussion} we discuss the origin
of charge dependent migration barrier and we compare our results with
experimental and theoretical works.

\section{Details of the simulation}
\label{detail}

Our simulations are performed using SIEST-A-RT, a method combining a
self-consistent density functional method (SIESTA)~\cite{3SAN97} with
the most recent version of activation relaxation technique (ART
nouveau)~\cite{3Malek_ART}. Integrating various empirical potentials,
ART nouveau was shown to sample efficiently the energy landscape of
amorphous semiconductors~\cite{3Valiq03}, glasses~\cite{3Malek_ART}, and
proteins~\cite{3WMD02}, for example. 

SIEST-A-RT was used for the first time to study vacancy diffusion in Si and
details of its implementation can be found in Ref.~\cite{3Elm04}. Forces and
energies are evaluated using density-functional theory (DFT) with the
local-density approximation (LDA) using standard norm-conserving
pseudo-potentials of Troullier-Martins~\cite{3Trou91} factorized in the
Kleiman-Baylander form~\cite{3Klei82}. Matrix elements are evaluated on a 3D
grid in real space. The one-particle problem is solved using linear
combination of pseudo-atomic orbitals (PAO) basis set of finite range. Here,
we use the extended local basis set~\cite{3Elm05}, which was shown to reproduce
closely the best solution while minimizing computational costs. Calculations
are performed on a 215-atom GaAs supercell sampled at the $\Gamma$ special
point. All relevant charge states (0, $-1$, $-2$ and $-3$) are fully relaxed
until the residual force falls below 0.002~eV/\AA~, then we proceed to search
for local diffusion pathways by assuming that the charge state of the defect
is preserved during any transformation. Unless mentioned, all events start
from fully relaxed gallium vacancy geometries.

In order to break the initial local symmetry, activated events are started by
displacing in a random direction a region of the cell centered around a chosen
atom in the first, second or fourth-shell around the vacancy, involving
typically between 10 and 30 atoms. The structure around the vacancy is then
deformed along this random direction until the lowest curvature, corresponding
to the lowest eigenvalue of the Hessian matrix, becomes negative, falling
below a preset threshold value. The system is then pushed along the
corresponding eigenvector, while minimizing the energy in the perpendicular
hyperplane, until the total force falls below 0.1~eV/\AA, indicating that the
transition state has been reached.

About 60 events were generated in total for all charge states, with 20 events
for jumps to the first nearest neighbor and 30 for jumps to the second
neighbor. In both cases the structure is deformed by at least 0.9~\AA~ before
a sufficiently negative eigenvalue appears. Diffusion to the fourth neighbor
was more difficult to complete (we generated 10 such events in total) and we set up a threshold displacement of 1.4~\AA~ to allow the configuration to escape from  the harmonic well.

\section{Results}
\label{results}
\begin{table*}
\caption{Nearest neighbor distances (in \AA ) in 111 direction relevant for
first nearest neighbor diffusion from the initial $\rightarrow$ to the
final state . Distances are calculated by taking the initial position
of $V_{Ga}$ as reference. The last column describes the geometry of  the final state.}
\begin{ruledtabular}
\begin{tabular} {lcccl}
& $V_{Ga}-As^{1st}$  &   $As^{1st}-Ga^{2nd}$ &$Ga^{2nd}-As^{3rd}$ &Final geometry \\
\\
\hline
  
 $V_{Ga}^{0}$ &2.08$\rightarrow$0.44 &2.43$\rightarrow$3.40 & $2.45\rightarrow$2.46  &$(As_{Ga} +V^{1st}_{As})^0$\\
 $V_{Ga}^{-1}$ &2.08$\rightarrow$0.90 &2.42$\rightarrow$2.65 & 2.45$\rightarrow$2.58 &$(V_{Ga} + I_{As} +V^{1st}_{As})^{-1}$\\
 $V_{Ga}^{-2}$ &2.06$\rightarrow$0.53 &2.41$\rightarrow$2.69 &2.45$\rightarrow$2.64  &$(As_{Ga} + V^{1st}_{As}
+I_{Ga} +V^{2nd}_{Ga})^{-2}$\\
\\
 $V_{Ga}^{-3}$ &\multicolumn{2}{l}{\it Does not diffuse to the first neighbor}\\
\end{tabular}
\end{ruledtabular}
\end{table*}

\subsection{Gallium vacancies at the initial state}

\begin{figure}[t]
\centerline{\includegraphics[viewport=20 90 400 400, width=7cm, angle=0]{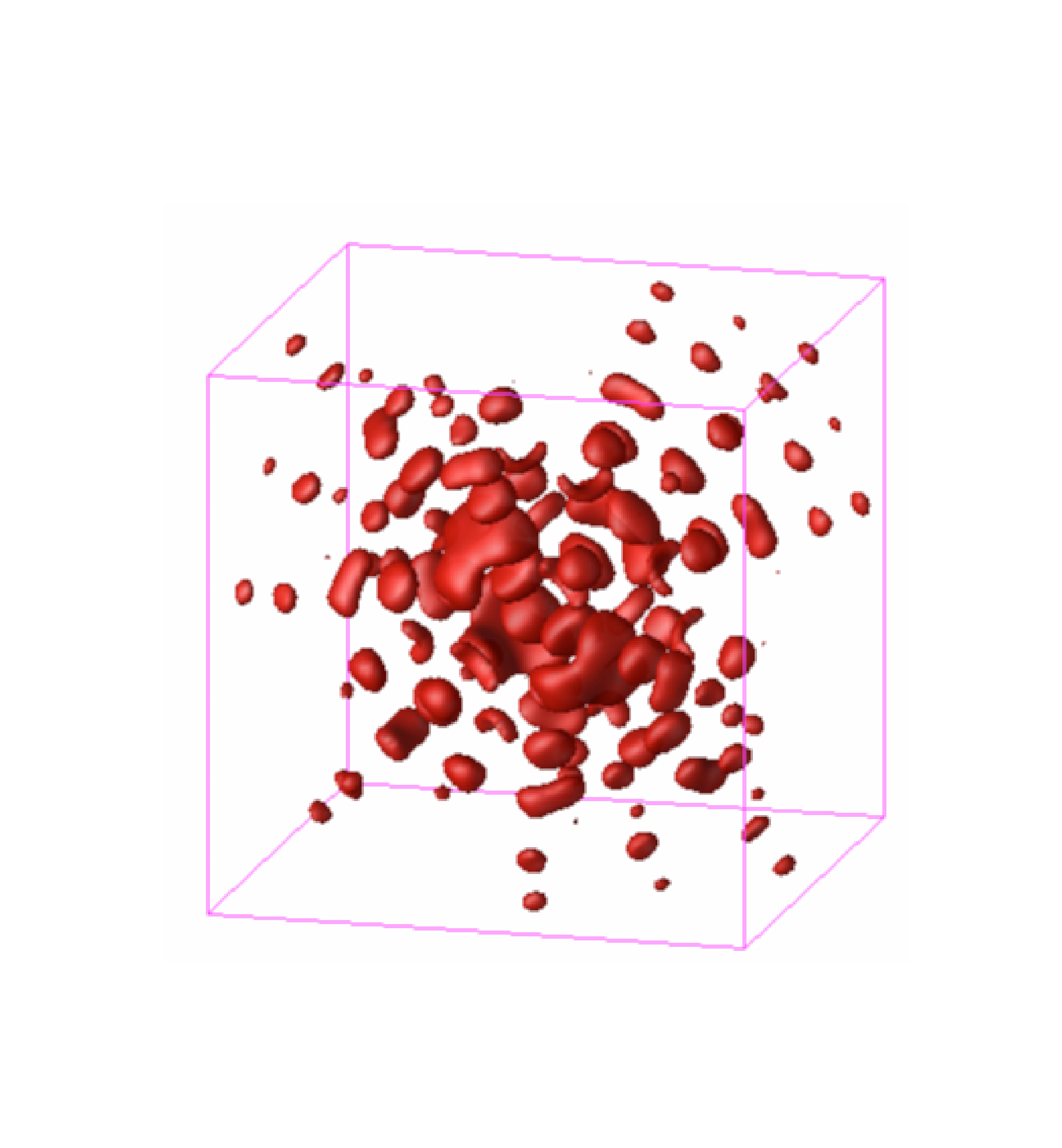}}
\caption{(Color online) Constant electron density   surface at 0.0004 electron/\AA$^3$ which shows the degree of localization of extra  $-3$ charge around a gallium vacancy in GaAs  located initially at the center of the  box.}
\label{fig:contour}
\end{figure}

 We first examine the
vacancy at the initial state before moving to the analysis of transition
states. The structure of $V_{Ga}$ for all charge states, for a number of basis
sets within SIESTA, is presented in Ref.~\cite{3Elm05} and agrees well with
other simulations and experiments. Relaxed at its energy minimum, $V_{Ga}$
conserves the $T_d$ symmetry for all charges, while the open volume associated
with the vacancy decreases with increasing charge. Spin-polarized LDA
relaxations for $V_{Ga}$ also lead to a $T_d$ symmetry, confirming that the
symmetry conservation of $V_{Ga}^q$ is not a drawback of LDA but a behavior
proper to cation vacancies, as pointed out previously by Chadi~\cite{3Cha03}.
This behavior has been further confirmed by a recent calculation~\cite{3Mak06}
on $V_{Ga}$ in GaAs using two-component density-functional theory-LDA applied
to a cubic 64 atom supercell together with a $4^3$ k-point mesh Brillouin zone
sampling .

A detailed analysis of the charge densities plots (see Fig.~\ref{fig:contour})
and the Mulliken populations reveals that the electrons added to the neutral
vacancy are delocalized and distribute themselves on the 111 axes passing by
the As dangling bonds in a similar way for all studied charge states. Less
than 4\% of the charge is localized on the four As neighboring the vacancy
($As^{1st}$) while the remaining 96 \% is spread over more distant neighbors
on the 111 axes. This suggests that these predominantly covalent As$-$Ga bonds
are progressively stiffened upon the addition of extra electrons, thus making
them more difficult to break compared to the remaining bonds and affecting directly the height of the diffusion barriers.

\subsection{Diffusion path to first  neighbor}

The activated events we generated using SIEST-A-RT show that diffusion to the
first nearest neighbor is not possible for all sequential charge states, contrary to what was
proposed by Van Vechten~\cite{3Vec84} nor impossible, in contradiction with what was found by Bockstedte and
Scheffler~\cite{3Boc96}. We find rather that diffusion to the first nearest
neighbor is very much charge dependent.

For $q=0$ the first neighbor of the vacancy ($As^{1st}$) diffuses, along the
111 direction, toward the vacant site via a split vacancy configuration by
optimizing its bonds with the close neighbors. This mechanism is similar to
the diffusion mechanism of a neutral silicon vacancy in
silicon~\cite{3Blo93,3Elm04,3Dab00}. First, each back bond of the diffusing atom
($As^{1st}-Ga^{2nd}$) stretches until it breaks during the migration of
$As^{1st}$ toward the vacancy. $As^{1st}$ proceeds in its migration until a
metastable vacancy-antisite structure ($As_{Ga} +V^{1st}_{As}$) forms. The
$As^{1st}-Ga^{2nd}$ bond evolves from 2.43~\AA\ at the initial state to
3.40~\AA\ at the final state. This metastable complex is 0.67~eV higher than
$V_{Ga}$ and the recorded migration barrier for the first neighbor diffusion
is 0.84~eV. The next first neighbor jump ($V^{1st}_{As}\rightarrow Ga^{2nd}$)
leading to the formation of $As_{Ga} + Ga^{1st}_{As} +V^{2nd}_{Ga}$ complex is
also possible by crossing a barrier of 1.55~eV.  Starting from this last complex we did not find any mechanism for $V^{2nd}_{Ga}$ to diffuse further to the first  neighbor,  suggesting that this jump is unfavorable.

For $q=-1$ the vacancy follows the same path to the saddle point as for $q=0$:
it crosses a barrier of 0.9~eV passing by a split vacancy site, but
$As^{1st}-Ga^{2nd}$ bond is less stretched. $As^{1st}$ relaxes then in a split
vacancy site, forming a $(V_{Ga} + I_{As} +V^{1st}_{As})^{-1}$ complex.
$As^{1st}-V_{Ga}$ distance is reduced to half (2.08~\AA\
$\rightarrow$0.90~\AA) suggesting that $As^{1st}$ is at half way between the
two vacancies. This metastable configuration is 0.81~eV higher than the
initial minimum. We confirmed that this final metastable state is not only a local minimum
along the diffusion path to the first neighbor by starting   from an ideal  vacancy-antisite complex  $(As_{Ga} +V^{1st}_{As})^{-1}$,  then relaxing  until the residual force becomes lower than 0.002~eV/\AA. This vacancy-antisite structure is found to be unstable in $-$1 charge state because $As^{1st}$ leaves the ideal antisite structure and prefers to relax at  a metastable state half way between the two vacancies. 

For $q=-2$, we find only a collective motion of a $As^{1st}-Ga^{2nd}$ pair
toward the vacancy along 111 direction. $As^{1st}-Ga^{2nd}$ bond stretches
slightly, while $As^{1st}$ approaches $V_{Ga}^{-2}$ as close as 0.53~\AA.
Consequently, $Ga^{2nd}$ is forced to stretch its back bonds and stabilizes in
an interstitial position. Finally, $As^{1st}$ atom occupies $V_{Ga}$ while
$Ga^{2nd}$ is located at a split interstitial position between $V^{1st}_{As}$
and $V^{2nd}_{Ga}$. The resulting metastable complex $(As_{Ga} + V^{1st}_{As}
+I_{Ga} +V^{2nd}_{Ga})^{-2}$ is 1.74~eV higher in energy than $V_{Ga}^{-2}$
and can be obtained by crossing a barrier of 1.86~eV. Relevant distances for
successful first nearest neighbor diffusion in $q = 0, -1, -2$ are summarized
in Table ~\ref{tab:distance}.

For $q=-3$, all attempts for first nearest neighbor diffusion failed
and the configuration always relaxes back to the initial minimum.
Even when forcing the jump by using the transition state at the neutral charge
state as starting point for a convergence of the $V_{Ga}^{-3}$ to its saddle
point, the vacancy systematically returns to its original state.

\subsection{Diffusion path  to fourth neighbor}
\begin{figure}[t]
\centerline{\rotatebox{0}{\includegraphics[viewport=90 90 450 450, width=7cm]{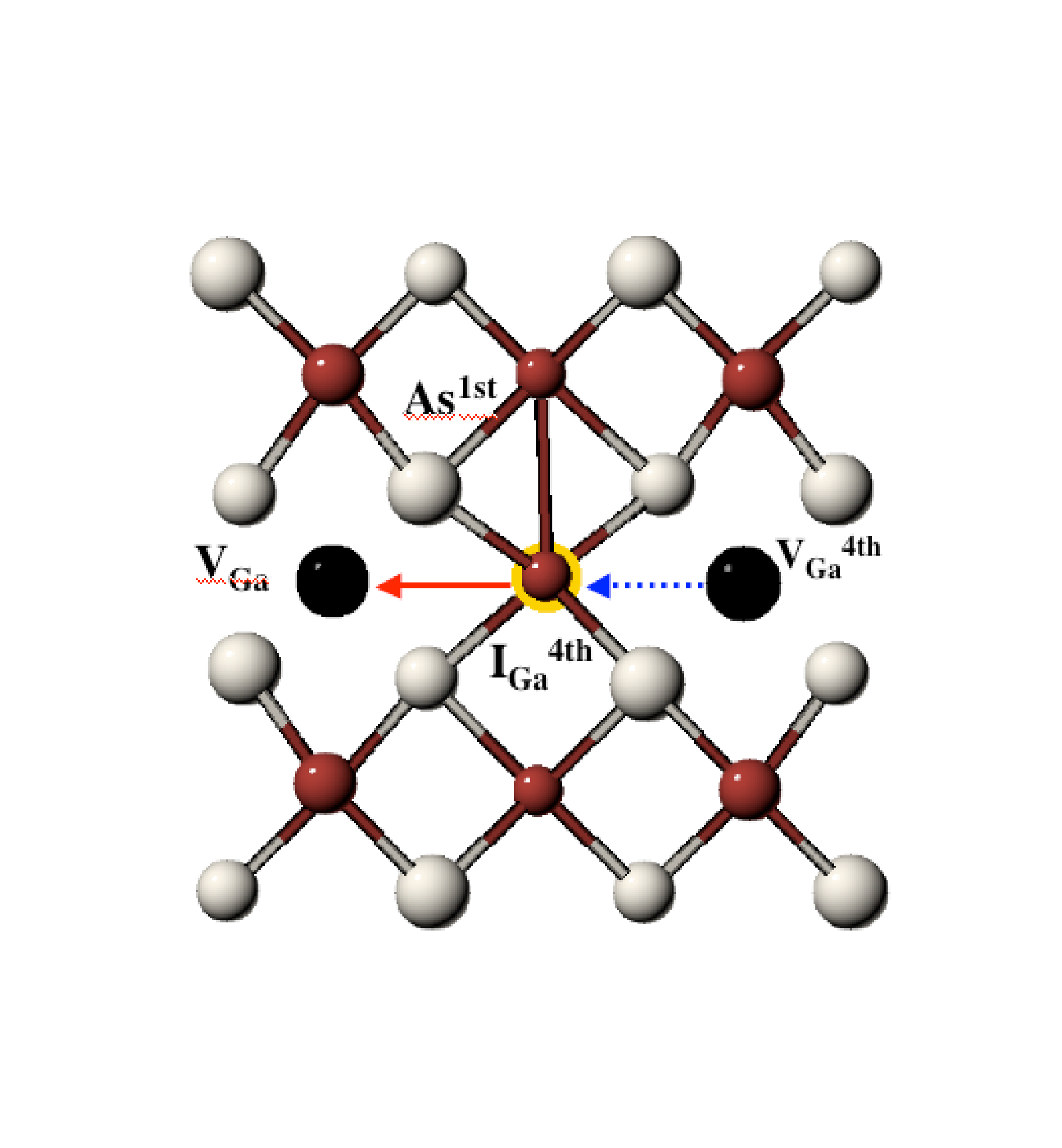}}}
\caption{(Color online) Migration path of $V_{Ga}^q$ to the fourth neighbor along (100) direction (see the text). }
\label{fig:Fourth}
\end{figure}

The situation is almost opposite for the diffusion to the fourth neighbour:
all negatively charged vacancies succeed in diffusing along this pathway,
only the neutral vacancy refuses to go this way. 

The fourth neighbor of the vacancy ($Ga^{4th}$) approaches the
interstitial region near the vacancy by diffusing along the 100
direction. Figure ~\ref{fig:Fourth} illustrates this configuration:
the dashed arrow shows the direction of the jump from the initial
state to the saddle point, while the full arrow shows the path from
the saddle to the final. $Ga^{4th}$ diffuses to an unstable
interstitial position close to the vacancy. In the cubic zincblende
structure of GaAs, with lattice constant $a$ and a vacancy located
initially at (0,0,0), the fourth neighbor diffuses first from (a,0,0)
to the interstitial position at (a/2, 0,0). The transition state of
this mechanism can be described as a gallium interstitial placed
between two distant vacancies ($V_{Ga}-I_{Ga}^{4th}-V_{Ga}^{4th}$).
This interstitial atom is not a direct neighbor of either of the two
vacancies because the $Ga^{4th}-V_{Ga}$ distance shortens from
5.6~\AA\ at the initial minimum to 2.88~\AA\ at the transition
state. This diffusion mechanism require an elevated barrier (around
4.24~eV) since the $Ga^{4th}$ needs to break two bonds initially with
$As^{5th}$ farthest away from the vacancy and to twist the remaining
$As^{3rd}$ bonds. Once it arrives at the saddle point two new bonds
with $As^{1st}$ are formed. It is interesting to note,
as is shown in Table~\ref{tab:barrier} that the energy barrier is
almost independent of the charge for the defects that manage to
diffuse to the fourth neighbour; this weak dependence can probably be
attributed to the constant electronic charge distribution along 100
direction for different negative charging.

\subsection{Diffusion path to  second  neighbor}
\begin{figure}[t]
 \centerline{\rotatebox{0}{\includegraphics[viewport=90 90 450 450, width=7cm]{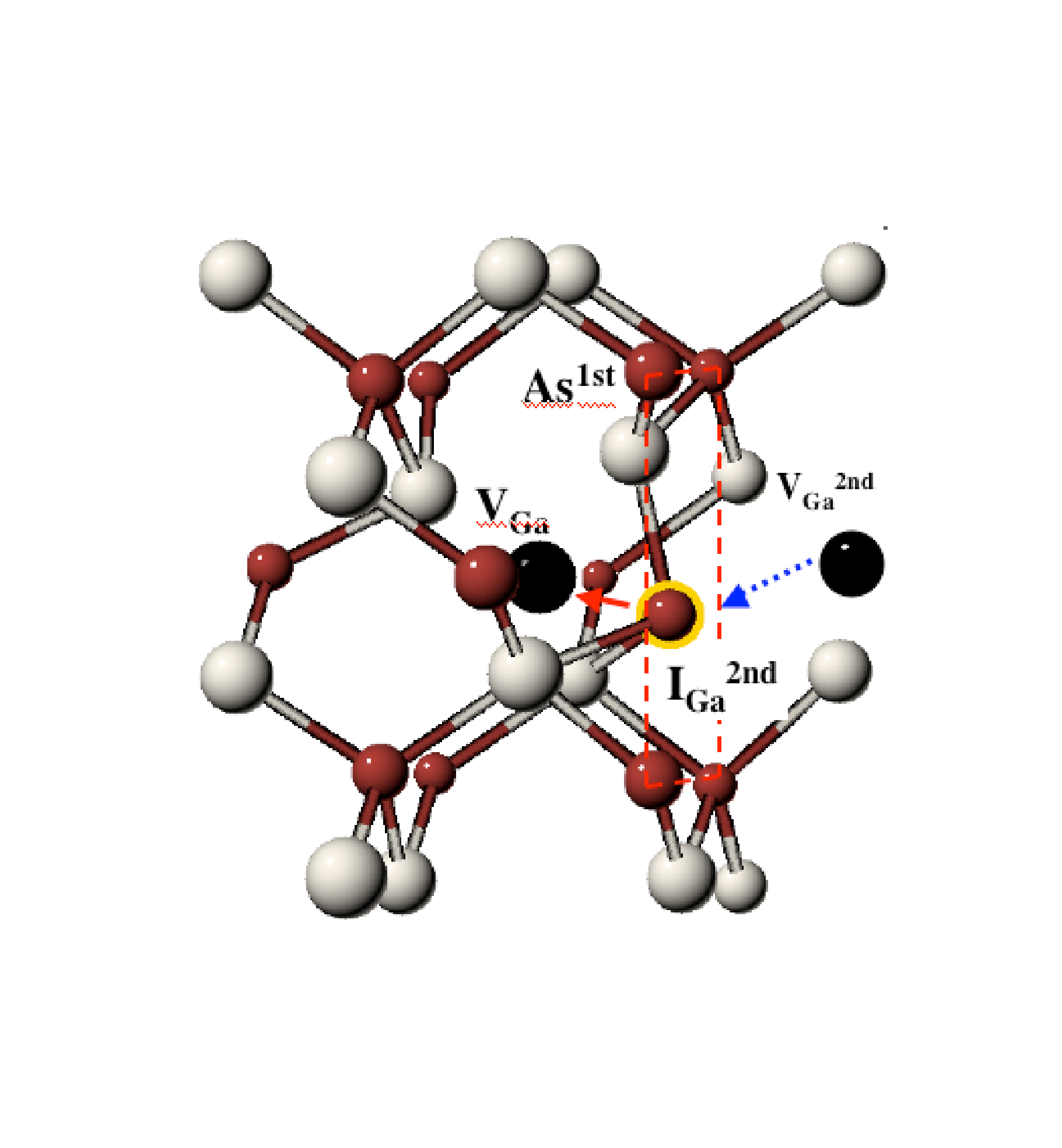}}}
 \caption{(Color online) Diffusion to second neighbor by plane-passing mechanism
   (refer to the text).}
 \label{fig:PP}
\end{figure}
By exploring the energy landscape of the system, we find that vacancy
migration to the second neighbor occurs by two mechanisms. The
simplest diffusion pathway is already well known~\cite{3Boc96} and is
considered to mediate self-diffusion in binary semiconductors, while
the second pathway, which has not been reported, to our knowledge, for
this system, is more complex and involves the correlated motion of
many atoms neighbouring the vacancy.

\subsubsection{Plane-passing mechanism}
\begin{figure}
  \centerline{\includegraphics[viewport= 180 100 450 750, width=5cm,angle=-0]{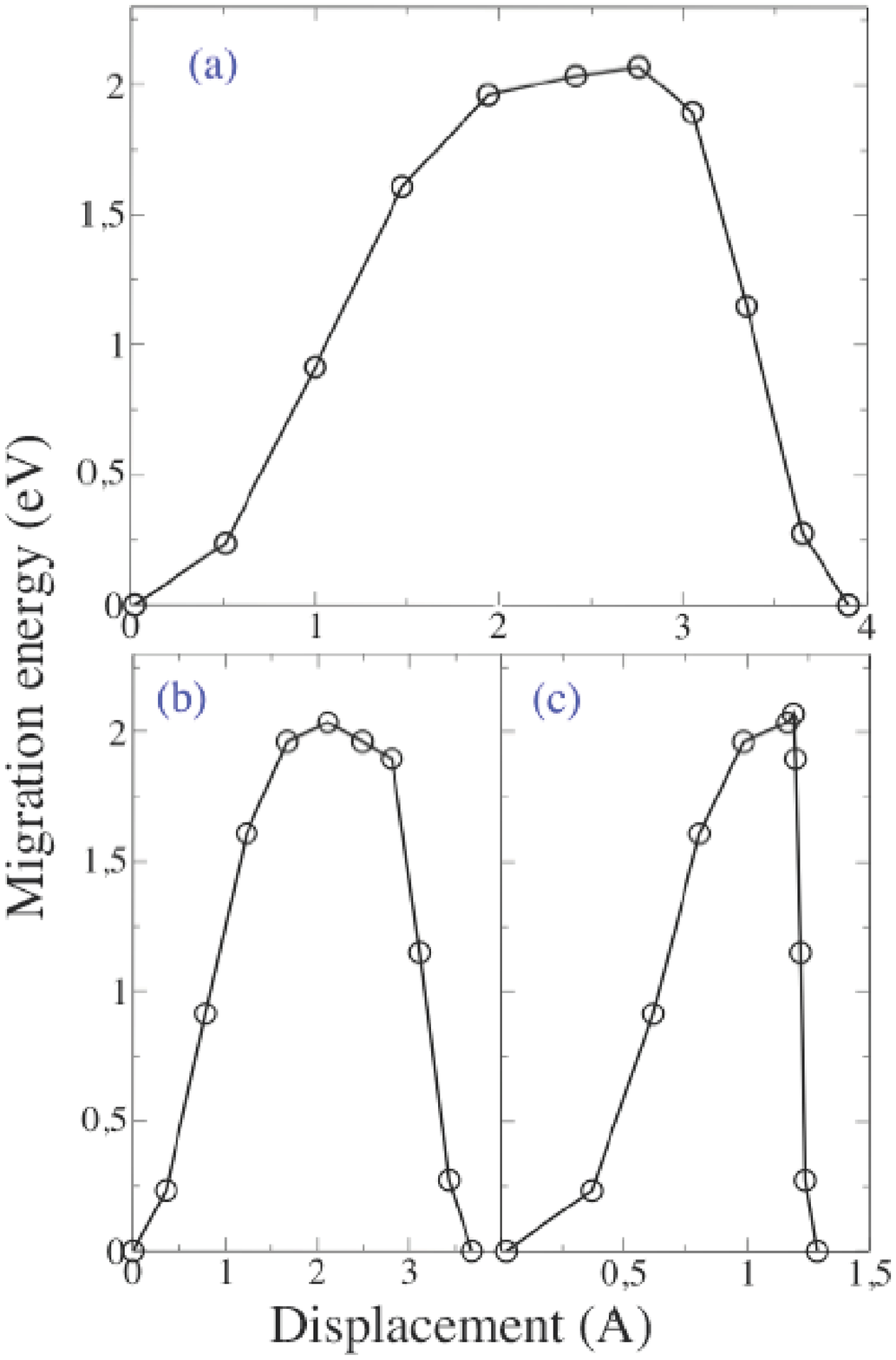}}
 \caption{(a) Migration trajectory to the second neighbor by the
 plane-passing mechanism for $-3$ charge state. 11 images (circles)
 are relaxed with CI-NEB method until  a  force tolerance of 1~eV/\AA~ is reached, lines are guide to the
 eye. The contributions from different moving atoms are decomposed in
 two: (b) The path followed by the diffusing atom; (c) Arrangement of
 the other atoms around the defect. }
  \label{fig:PP_path}
\end{figure}

The most intuitive diffusion pathway, which we call the {\it plane-passing
mechanism}, brings one $Ga^{\mathrm{2nd}}$ to the interstitial region, joining
the diffusing atom and the vacancy, along the 110 direction. The
diffusing $Ga^{\mathrm{2nd}}$ atom must go through the diffusion plane
perpendicular to the 110 direction during its way to the vacancy site
causing three $As^{1st}$ atoms to move away from the vacancy and
opening the cage. The transition state of this mechanism, first
identified by Bockstedte and Scheffler~\cite{3Boc96}, can be described
as a gallium interstitial placed between two close vacancies ($V_{Ga}
+ I_{Ga}^{2nd} + V_{Ga}^{2nd}$). Figure ~\ref{fig:PP} illustrates this
configuration: the dashed arrow shows the direction of the jump from
the initial state to the saddle point, while the full arrow shows the
path from the saddle to the final; the diffusion plane formed by the
second nearest neighbors of the vacancy is also shown.

From Table~\ref{tab:distance}, it is clear that, as with the other mechanisms 
to first and fourth neighbors, the structural
details of the jumps are charge-dependent. The position of the
transition state, in particular, changes by 20 \% as the vacancy goes
from neutral to a charge of $-3$. As can be expected, the displacement
of the unstable interstitial position as the diffusing atom becomes a
nearest neighbour of the vacancy has noticeable impact on the energy
barrier, which goes from  1.7~eV for neutral and $-$1 charge state
to 1.84 and 2.0~eV for $-$2 and $-$3 charges respectively.
\begin{figure}[t]
  \centerline{\includegraphics[viewport= 150 150 500 700, width=7cm]{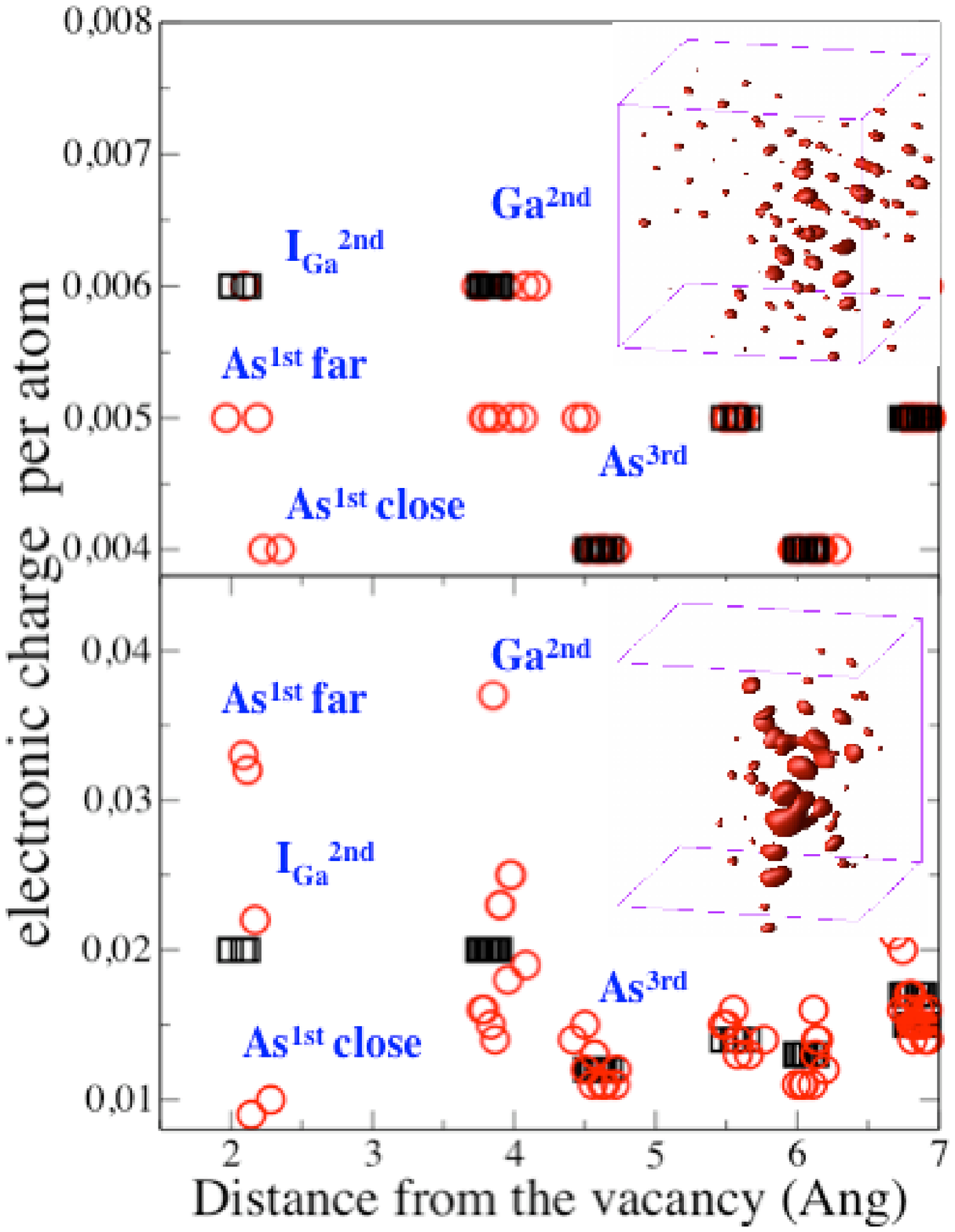}}
  \caption{(Color online)  Mulliken population analysis of charge
    distribution around the vacancy at initial (square) and the saddle
    point (circle) of plane-passing mechanism.  Shown the degree of
    localization of extra electrons at typical distances from the
    vacancy for charge $-$1 (top panel), and $-$3 (low panel). The
    insets show the constant electron density surface at 0.00018 and
    0.0008 electron/~\AA$^3$ near $V_{Ga}^{-1}$ and $V_{Ga}^{-3 }$
    respectively. }
  \label{fig:Mulli_PP}
\end{figure}

\begin{table}
   \caption{Evolution of the distance (in~\AA) between the initial
   position of the vacancy and the diffusing atom
   ($V_{Ga}^{q}-Ga^{2nd}$) in the initial state and the transition
   state for both diffusion mechanisms to the second neighbour.}
\label{tab:distance}
\begin{ruledtabular}
\begin{tabular} {l|cc|cc}
& \multicolumn{2}{c}{\bf Plane-passing}  
& \multicolumn{2}{c}{\bf Cluster-assisted}   \\
 \hline
  & Initial & Saddle & Initial & Saddle \\
\\
\hline
 $V_{Ga}^{0}$ &3.89 &1.89 &\\
  $V_{Ga}^{-1}$ &3.88 & 2.09 &3.88 & 2.5 \\
  $V_{Ga}^{-2}$ &3.82 & 2.19 &3.82 & 2.78 \\
 $V_{Ga}^{-3}$ &3.80 & 2.35 &3.80 & 2.89 \\
\end{tabular}
\end{ruledtabular}
\end{table}

More precisely, for $V_{Ga}^0$ and $V_{Ga}^{-1}$, the saddle point is
a site close the hexagonal interstitial, while it is closer to the
tetrahedral interstitial configuration for the more negative
$V_{Ga}^{-2}$ and $V_{Ga}^{-3}$. At the transition state, as the
charge increases, the displacement of the gallium atom out of the plane
becomes less pronounced; for the $-3$ charge state, it almost vanishes,
leaving the moving atom on the plane.

This can be seen by looking at the the full migration trajectory for
$-$3 charge state in Figure~\ref{fig:PP_path}. This path is generated
by initially interpolating between the ART-generated initial, saddle
and final states, generating 11 images.  These image configurations
are then relaxed using the CI-nudged-elastic-band method~\cite{3UBER00}
until the total force becomes lower than 1eV/\AA~. After decomposing
the total path into contributions coming from different moving atoms one
can notice that the path followed by the diffusing atom is nearly
symmetric, while most asymmetry in the total path comes from
arrangement of the other atoms around the defect.

Differences in the migration barriers are mainly due to the diffusion
of the electronic charge during the jump. For the singly negative
vacancy, Mulliken population analysis at the saddle point (top panel
in Figure~\ref{fig:Mulli_PP}) is compared to the initial
configuration. It shows the diffusion of the electronic charge during
the migration of the $Ga^{2nd}$ atom. The electronic charge at
$Ga^{2nd}$ diffuses with it to the hexagonal interstitial site thus
saturating partially the $As^{1st}$ dangling bonds. The charge is
consequently suppressed from the neighborhood of the diffusing atom
($I_{Ga}^{2nd}$) and spreads uniformly over more distant shells as can
be seen from the 3D charge density (top inset of
Figure~\ref{fig:Mulli_PP}).

For $V_{Ga}^{-2, -3}$ the diffusing atom is less engaged toward the
vacancy: circles in the lower panel in Figure~\ref{fig:Mulli_PP} are
shifted to the right as a signature of volume opening that affects the
second neighbor of the vacancy as well. However, the 3D charge
densities for $V_{Ga}^{-2}$ and $V_{Ga}^{-3}$ are different from
$V_{Ga}^{-1}$. After the jump, the charge becomes highly localized
around the dangling bonds belonging to As atoms farthest away from
$I_{Ga}^{2nd}$ labeled $As^{1st}_{far}$. Some of the charge carried by
two closer $As^{1st}_{close}$ to $I_{Ga}^{2nd}$ get transferred to the
other $As^{1st}_{far}$; these have non saturated bonds that are still
pending.

\subsubsection{Cluster-assisted mechanism }

\begin{figure}
  \centerline{\rotatebox{0}{\includegraphics[viewport=90 90 450 450,width=7.4cm]{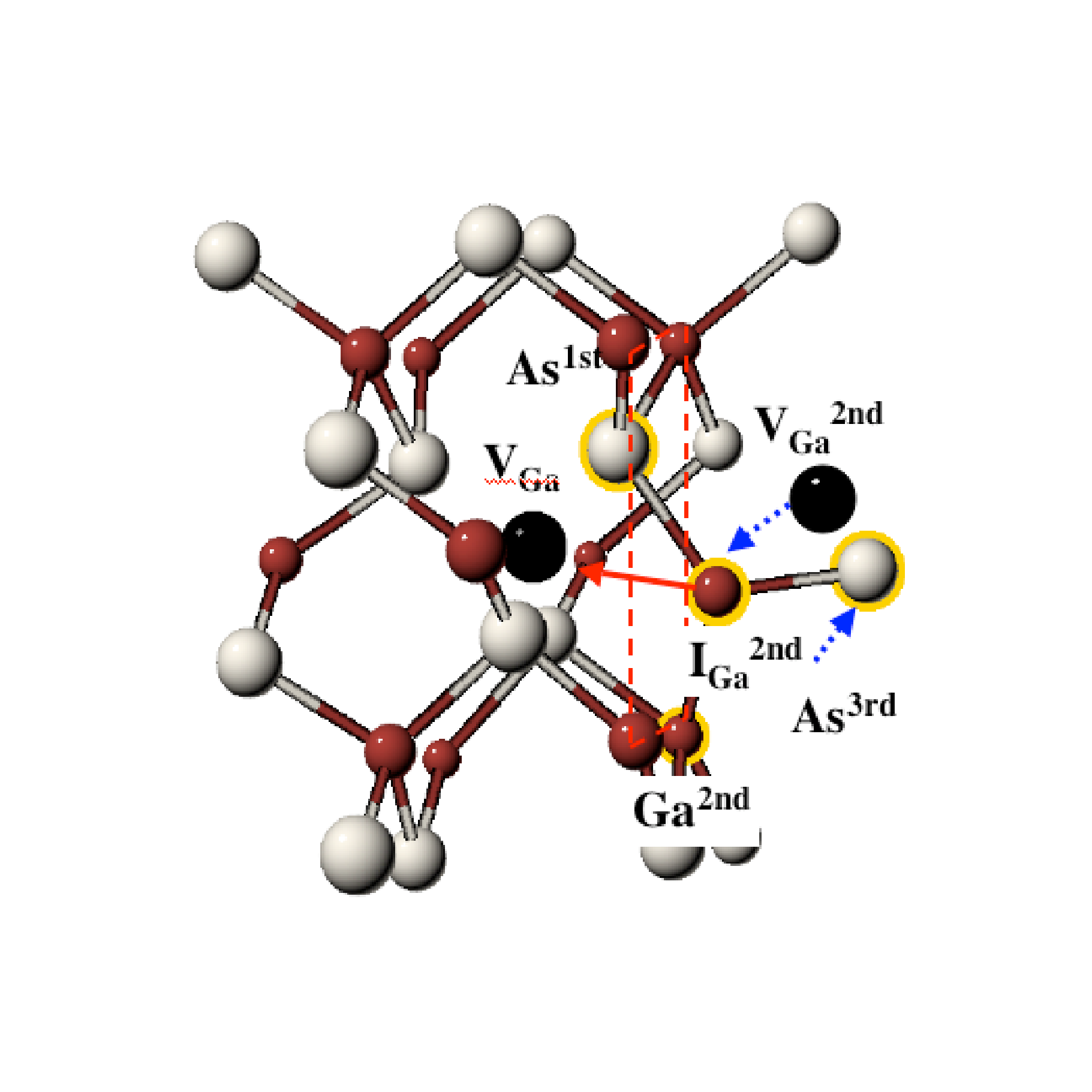}}}
  \caption{ (Color online) Diffusion to second neighbor by
  cluster-assisted mechanism (refer to the text).}
  \label{fig:PA}
\end{figure}

Negatively charged $V_{Ga}$ can also diffuse on the Ga sublattice
through a mechanism that we call the {\it cluster-assisted path}. As far as we know,
 this diffusion pathway had not been reported until now.  Instead of crossing directly the diffusion plane, $Ga^{2nd}$
approaches the interstitial region far from the plane, being assisted
by two As atoms--- respectively first ($As^{1st}$) and third neighbors
($As^{3rd}$) of the vacancy--- and one gallium atom second neighbor of
the vacancy $Ga^{2nd}$. From the initial to the transition state
(dashed arrows on Figure~\ref{fig:PA}), an incomplete bond exchange
mechanism of type Wooten-Winer-Weaire~\cite{3WWW85} ---which we also
find in Si--- occurs between $Ga^{2nd}$ and $As^{3rd}$, then
$Ga^{2nd}$ is pushed into the interstitial region.

\begin{figure}
  \centerline{\includegraphics[viewport= 180 100 450 750, width=5cm,angle=-0]{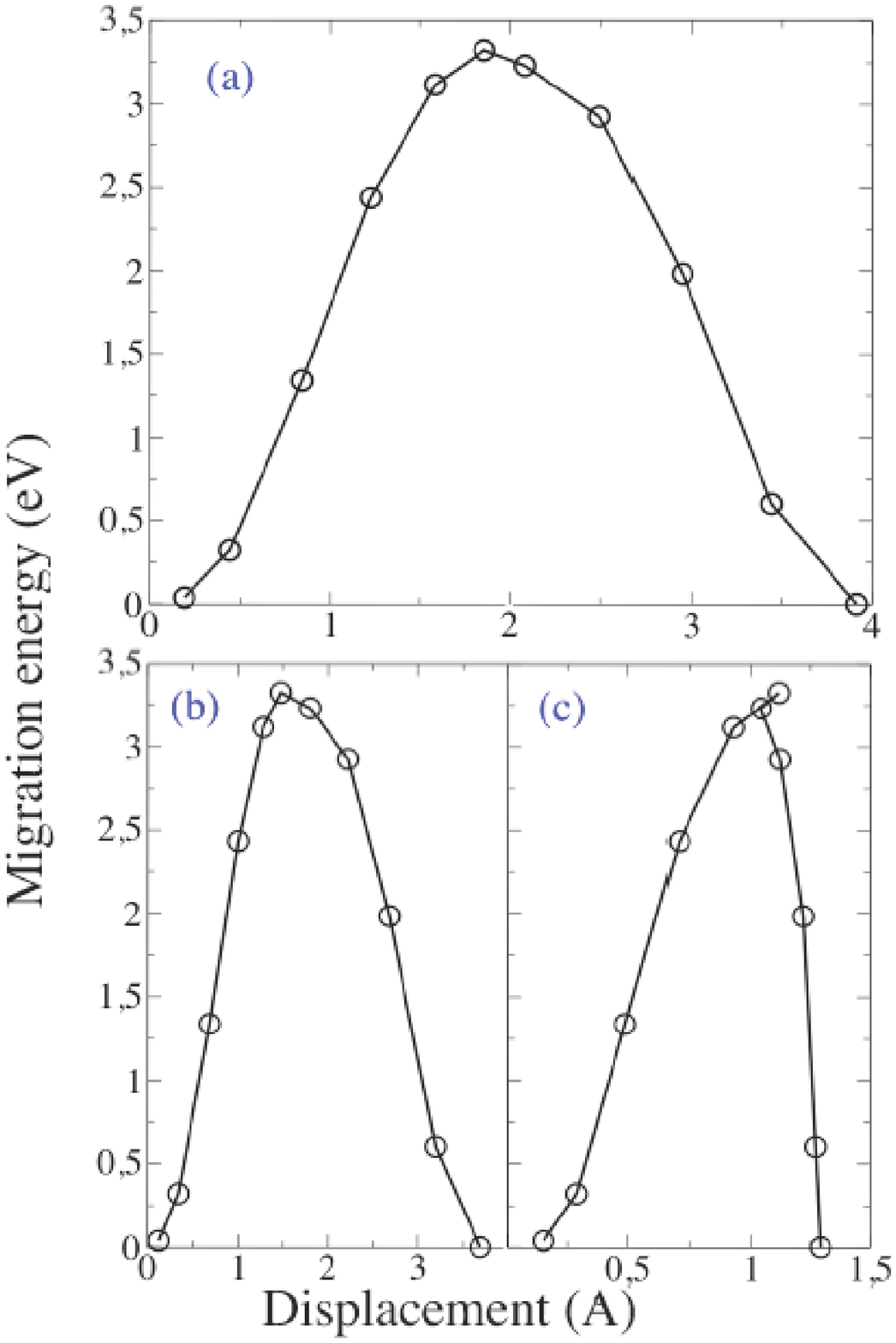}}
 \caption[Migration trajectory to the second neighbor by the
 cluster-assisted mechanism for $-3$ charge state.]{(a) Migration trajectory to the second neighbor by the
 cluster-assisted mechanism for $-3$ charge state. 11 images (circles)
 are relaxed with CI-NEB method until  a  force tolerance of 1~eV/\AA~ is reached, lines are guide to the
 eye. The contributions from different moving atoms are decomposed in
 two: (b) The path followed by the diffusing atom; (c) Arrangement of
 the other atoms around the defect. }
  \label{fig:CA_path}
\end{figure}

The cluster formed by $As^{1st} +I_{Ga}^{2nd}+As^{3rd} +Ga^{2nd}$
plays the major role for diffusion since the bond distances
$I_{Ga}^{2nd}-As^{1st}$ and $I_{Ga}^{2nd}-As^{3rd}$ remain unchanged
($~$2.4~\AA) regardless of the charge state, while the
$V_{Ga}^q-I_{Ga}^{2nd}$ distance increases by adding extra electrons
(see Table~\ref{tab:distance}). This assumes that the whole cluster
becomes less engaged toward the vacancy when passing from $-1$ to $-3$
charge states. During the relaxation from the saddle point to the
final state (full arrow on Figure~\ref{fig:PA}) only the gallium atom
at the interstitial position ($I_{Ga}^{2nd}$) continues its motion
toward the vacancy leaving the remaining constituents of the cluster
close to their initial positions. This is confirmed by looking at the
full migration trajectory for $-$3 charge state for the
cluster-assisted mechanism plotted in Figure~\ref{fig:CA_path}. From
the initial to the saddle point many atoms are experiencing
rearrangements and displacements, but the main contribution comes from
the diffusing atom which is less engaged toward the vacancy than for
plane-passing mechanism, while during the relaxation from the saddle
point to the final state the main contribution to the total path
comes from the relaxation of $I_{Ga}^{2nd}$ atom
(Figure~\ref{fig:CA_path}(b)).

Mulliken population analysis shows that the electronic charge diffusion
for cluster-assisted mechanism is completely different from the
plane-passing mechanism. By increasing the charge state of the vacancy
from $-1$ to $-3$, positions of the cluster atoms are all shifted to
the right confirming that the whole cluster is experiencing a
displacement far away from the vacancy. In addition, the electronic
charge becomes more and more concentrated around the cluster as the
charge on the defect increases. Consequently, the migration barrier
increases in significant way with the charge state reaching 2.44, 2.89
to 3.49~eV for $-$1, $-$2, $-$3 charges respectively.

 \begin{table}[t]
   \caption{Calculated diffusion barriers (in eV) for
   $V_{Ga}^{0,-1,-2,-3}$ in GaAs for all possible migration paths
   identified. The empty cell means that the migration was not
   possible with this mechanism.}
\label{tab:barrier}
\begin{ruledtabular}
\begin{tabular} {lcccc}
& $V_{Ga}^{0}$ & $V_{Ga}^{-1}$ & $V_{Ga}^{-2}$& $V_{Ga}^{-3}$   \\
\\
\hline

{\bf First neighbor} &0.84 &0.90 &1.86 &\\
{\bf Plane-passing}  &1.7 &1.7 &1.85 &2.0  \\
 {\bf Cluster-assisted}    &  &2.44  &2.89&3.24\\ 
 {\bf Fourth neighbor} &    &4.24 &4.24 &4.3\\
 \end{tabular}
\end{ruledtabular}
\end{table}

No cluster-assisted path  was found  at the neutral state:  we tried to
generate this event by  starting  directly from   the transition state at 
$-1$ charge, but, the neutral  vacancy relaxes back to the initial minimum  
rather than diffuse to the second neighbor site, demonstrating that this 
path is impossible at the neutral charge state.

\subsection{Summary of the results}

The summary of the calculated migration barriers is presented in 
Table~\ref{tab:barrier}: First neighbor diffusion has the lowest
barrier for 0 and $-1$ charges, but cannot be considered to be
dominant for $V_{Ga}$ since the complete diffusion process is
impossible to achieve even for the neutral vacancy.  However, this
mechanism induces the formation of defects belonging to the $As_{Ga}$
family of great importance since they are responsible of EL type
defects in GaAs (see the discussion below).

Self-diffusion will rather be dominated by  jumps to the same
sublattice.  Gallium vacancies can diffuse to the second neighbour
through the plane-passing mechanism for all charge states by crossing
barriers lower than 2~eV. The impact of the charge state on the
trajectory is reflected on the charge-dependent barrier-height energy,
for example, the barrier height for $V_{Ga}^{-3}$ is 18~\% higher than
for $V_{Ga}^{0}$, making the crossing 30 times less likely at
1000~K. The simplicity of this path as well as the relatively low
migration barrier make the plane-passing mechanism the potential
candidate to mediate vacancy self-diffusion in GaAs.

Negatively charged vacancies might also follow the cluster-assisted
pathway to diffuse to the second neighbor or diffuse directly to the
fourth neighbor with energy-barrier which is, by about 50~\% and
100~\% respectively, higher than that of the plane-passing
mechanism. Although these two mechanisms have elevated barriers, their
existence is very interesting by itself  as it shows the
underestimated richness of self-diffusion phenomena in bulk
semiconductors.

\section{Discussion and comparisons}
\label{discussion}

\subsection{Self-diffusion to first-neighboring sites}

In a binary lattice, vacancy hops to the first-neighbor sites require
a complex sequence of moves to preserve chemical ordering on the long
run. In 1984, Van Vechten~\cite{3Vec84} proposed a model leading to
vacancy diffusion to the second neighbor in GaAs.  His model describes a
11 first-neighbor hop process on a six membered ring and is based on two
assumptions: (i) jumps to the first neighbors are always possible, and (ii)
vacancy-antisite complexes are always stable. During successive hops
the vacancy should proceed leaving behind a chain of antisites, which
is energetically unfavorable. This is avoided by the vacancy passing
twice through the same sixfold ring. At the final stage all the
antisite defects are removed. Otherwise such a mechanism would create
an unfavorable excess of antisite defects beyond the equilibrium
concentration.  Bockstedte and Scheffler~\cite{3Boc96} studied the
validity of these assumptions ---possible diffusion via first neighbor
hops--- using the drag method and LDA. They found that $As_{Ga}
+V^{1st}_{As}$ is metastable in the neutral state while it is unstable
for other negative charge states for a 64-GaAs atoms system. They
concluded that diffusion by first nearest neighbor hops was
impossible. 

Our study shows that the complex is rather metastable in $-$1 charge
state too. The first nearest neighbor diffusion of $V_{Ga}$ in $-$1
and $-2$ charges is found to be possible via {\it deformed} structures
belonging to $As_{Ga}$ family. While $V_{Ga}$ diffusion to the first
neighbor lead to a symmetric $As_{Ga}-V_{As}$ complex in the neutral
state, this configuration is distorted in the case of $-$1 charge
states.  In the deformed structure for $-$1 charge state, As atom
do not occupy the tetrahedral gallium vacancy site, it is rather is
displaced from the Ga vacancy site by about 0.9 eV toward
$V_{As}^{1st}$. Interestingly, this deformed structure is no longer
metastable for $-2$, the transition state originates from a $As_{Ga}
+Ga_{As} + V_{Ga}$ complex, whith $Ga_{As}$ pushed toward $V_{Ga}$ and
occupying a split vacancy site.

We attribute this charge-dependent first-neighbor diffusion to two
competing factors: (1) the progressive increase in the strength of
$As^{1st}-Ga^{2nd}$ bonds as electrons are added make them more
difficult to break, and (2) the electron density in the region
surrounding the vacancy ---especially on As dangling bonds---
increases as electrons are added to the relaxed system. Thus, Coulomb
repulsion between three $As^{1st}$ atoms and the diffusing $As^{1st}$
atom can become so strong that this atom cannot approach further the
vacant site. This picture causes the $As^{1st}$ atom to relax at the
split vacancy configuration in $-$1 charge state. If the system is
charged $-$2, the $As^{1st}-Ga^{2nd}$ bond is so strong that
$As^{1st}$ pulls $Ga^{2nd}$ with it during its diffusion to the saddle
point position leaving behind $V_{As}^{1st}$ and $V_{Ga}^{2nd}$. The
electronic density is partially transferred from the initial vacancy
to these two new vacancies allowing $As^{1st}$ to relax on the initial
vacant site and leaving the $Ga^{2nd}$ atom stacked between
$V_{As}^{1st}$ and $V_{Ga}^{2nd}$.

Similarly deformed structures for negatively charged defects have been
recently reported in the litterature.  Chadi~\cite{3Cha03_2} found that
the isolated As antisite structure ($As_{Ga}$ is generally accepted to
be the basic structure of EL2 defect) exists in charge states $+$2,
$+$1 and 0 occupying the tetrahedrally symmetric position and in
$-$1,$-$2 charge states when it undergoes a small displacement that
causes a deviation from $T_d$ symmetry.  The main difference between
different charge states is the degree of relaxation that the antisite
and its direct neighbor undergo.  This possible high negative charging
of $As_{Ga}$ inducing a structural relaxation is similar to the
deformed negatively charged $As_{Ga}-X$ observed in this work.

Moreover, the electronic structure of $As_{Ga}-V_{As}$ has been
studied by total-energy Green's-function calculations treating
many-body effects within LSDA-DF~\cite{3Ove05} and shows that the
electronic levels allow at most $-$1 charging for this complex which
is in agreement with our results.

On the experimental side, negatively charged $V_{As}-X$ complexes have
been recently detected by positron annihilation
experiments~\cite{3Bon05} between 20$-$330~K in weakly p-type GaAs
under arsenic rich condition.  The detection of $V_{As}-X$ complexes
in these samples was surprising and have never been reported before
since under these conditions~\cite{3Elm05} isolated arsenic vacancies
are unlikely to form.  $V_{Ga}$ diffusion to the first neighbor after
annealing could explain these finding assuming that the metastable
complex $As_{Ga}-V_{As}$ could be detected at sufficiently low
temperatures.

In addition, $As_{Ga}-V_{As}$ structure was proposed by Steinegger
{\it et al.}~\cite{3Ste01} to be a potential candidate for EL6 defect.
Measuring the relative concentration of EL6 by photon-induced
current-transient spectroscopy at room temperature, they observed an
increase of $As_{Ga}$ concentration with annealing time by roughly a
factor of 2 corresponding to a to decrease of $V_{Ga}$ concentration
from about $10^{16}$ to zero.  While the first nearest-neighbor
diffusion cannot play a role in self-diffusion, we suggest that it is
dominant in these conditions, leading to the transformation of a large
number of $V_{Ga}$ into $As_{Ga}-V_{As}$ complexes.

\subsection{Diffusion to second neighbor} 
 
Nevertheless, diffusion to the second neighbor is more interesting
since it conserves the equilibrium concentration of vacancies. The
plane-passing paths of $V_{Ga}$ simulated with SIEST-A-RT confirm that
the transition state does cross the diffusion plane in all charge
states. For all our simulations the $As^{1st}$ atom crosses the plane
before $Ga^{2nd}$ atom;  we did not record any event where
$As^{1st}$ crossed out the plane or where it was located on the plane
as suggested by Bockstedte and Scheffler~\cite{3Boc96}.  However, the
presence of the extra charge affects the diffusion trajectory by
increasing the distance between the moving atom ($Ga^{2nd}$) and the
three $As^{1st}$ dangling bonds. For $V_{Ga}^{-2}$ and $V_{Ga}^{-3}$
the charge on $As^{1st}$ is so strong that is scatters the moving atom
and pushes it away from the vacancy toward the plane. Consequently,
the distance between the plane and the atom is lowered as the charge
of the vacancy increases, it almost vanishes for $-$3 charge state.

The diffusion barrier of plane-passing mechanism is found to {\it
increase} moderately as extra electrons are added. Recent theoretical
works reported a charge-dependent migration to the second neighbor by
plane-passing mechanism for vacancies in SiC and GaN binary
semiconductors.  Bockstedte {\it et al.}~\cite{3Boc03} found that the
migration barrier for diffusion $V_{C}$ and $V_{Si}$ in 4C-Si {\it
decreases} when the vacancy charges are {\it increasing}
progressively. A similar trend was observed for $V_{N}$ in
GaN~\cite{3Lim04} and recently for $V_{Ga}$ in GaN~\cite{3Gan06}.  Our
most recent results with SIEST-A-RT concerning $V_{As}$ in
GaAs~\cite{3Elm06_2} suggest that the migration barrier by
plane-passing mechanism {\it decreases} by {\it increasing}
progressively vacancy charging. 
Thus, $V_{Ga}$ in GaAs shows an opposite  trend compared to previously
studied vacancies.  This trend was observed previously by Bockstedte
and Scheffler~\cite{3Boc96} as they found a migration barrier of 1.7~eV
for neutral and 1.9~eV for $-$3 charge state. This behavior cannot be
a drawback of SIEST-A-RT, since the total path relaxed using CI-NEBM
lead to the same barrier, it is rather correlated to the electronic
charge diffusion of $V_{Ga}$ in GaAs.

Experimentally, negatively charged gallium vacancy migration on GaAs(110)
surface have been found to be stimulated by the tip during STM
experiments~\cite{3Len96} only when {\it holes} are injected on the
surface reducing consequently  the negative charging of the vacancy. This supports our results suggesting that the migration barrier is {\it lowered} as the vacancy becomes less negatively charged. 

\subsection{Diffusion in experimental systems}

Bliss {\it et al.}~\cite{3Bli93} identified the migration barriers for
$V_{Ga}$ in LT-GaAs by positron annihilation technique. They found a
migration enthalpy of 1.5$\pm$0.3~eV for the vacancy diffusion to the
second neighbor and 1.1$\pm$0.3~eV for diffusion to the first
neighbor. Within the experimental error bars, these results agree with
our calculated barriers summarized in Table~\ref{tab:barrier}. Our
results are also consistent with the widely accepted value of
1.7-1.8~eV for GaAs vacancy migration in bulk and GaAs-based QW
materials~\cite{3Rou92, 3Geu99, 3Bli92, 3Lah96, 3Bal00}. This value has
been recently reported for inter-diffusion in InGaAs/GaAs quantum
dots~\cite{3Dji06} as well. Thus our calculated diffusion barriers are
accurate enough and could be extended to analyze experimental data in
bulk and even in nanostructured materials.

Our study can be useful for interpreting more accurately
experimental data, especially if taking into account the fact that
most experiments for gallium self-diffusion~\cite{3Bra99_2, 3Geb03} have
been performed in n-type or intrinsic material where negatively
charged vacancies are more abundant. The most relevant pathway for
self-diffusion is the plane-passing mechanism since it has the lowest
barrier for diffusion to the second neighbor, while at sufficiently
low temperatures diffusion to the first neighbor could be observed for
$-1$ and $-2$ charge states. By combining information coming from our
calculated first and second neighbor diffusion barrier it becomes
possible to identify the charge state of the diffusing vacancy in some
experiments. For example, in the experiment of Bliss {\it et
al.}~\cite{3Bli93},  $V_{Ga}$ diffuses to the first neighbor by crossing a barrier of  1.1$\pm$0.3~eV,  which is close to our calculated barrier of 0.84~eV for $V_{Ga}^{-1}$. In addition, measured diffusion barrier to the second neighbor  is 1.5$\pm$0.3~eV  which is comparable to  1.7~eV we calculated for $-$1 charge state. 
Thus, we can state that diffusing vacancies in this experiment are more probably singly negative  instead of the generally accepted triply negative.  Recent
research works support more and more the hypothesis of $-1$ charge
state. This was the case, for example, in a recent theoretical study
comparing simultaneously the relaxations and lifetimes obtained from
simulation and Doppler spectra of positron annihilation experiments, the results confirm  the possibility of $-$1 charge state for $V_{Ga}$ rather that the $-$3 charge state~\cite{3Mak06}.

\section{Conclusions}
\label{conclusion}

The SIEST-A-RT approach has been used to study various diffusion
pathways for charged gallium vacancies in GaAs demonstrating that
diffusion in bulk semiconductors is a rich and complex phenomenon
closely related to the charge state.  Novel diffusion pathways are
identified for negatively charged vacancies, showing that diffusion
process could be non intuitive and more complex than initially
thought.  $V_{Ga}$ can diffuse to the second neighbor using two
different mechanisms in addition to diffusion to the first and fourth
neighbor.  Diffusion to the second neighbor is possible by the {\it
plane-passing mechanism} at all charge states as well as by the newly
found {\it cluster-assisted mechanism} which becomes possible only in
the presence of negatively charged vacancies. In addition, gallium
vacancy diffusion to the first nearest neighbor is possible only for $
q=0,-1, -2 $ charges passing by regular or distorted $As_{Ga}-X$
defects. However this mechanism cannot be considered to be dominant
for $V_{Ga}$ since the complete diffusion process was impossible to
achieve.  Finally, the highest diffusion barrier was recorded for the
direct diffusion to the fourth neighbor along the 100 direction for
negatively charged vacancies.  All barriers for the migration pathways
of $V_{Ga}$ are found to increase with increasing the number of
electron, a behaviour opposite to what was recently found in the case
of vacancy-mediated self-diffusion in  SiC~\cite{3Boc03,3Rur03} and
GaN~\cite{3Lim04, 3Gan06}.

{\it Acknowledgements.} We thank Professor Michel C\^ot\'e and Dr. Vladimir
Timochevski for fruitful discussions. NM acknowledges partial support
from FQRNT (Qu\'ebec), NSERC (Canada) and the Canada Research Chair program. We
thank the RQCHP for generous allocation of computer time.

\end{document}